\documentclass[12pt,reqno,twoside]{amsart}%
\usepackage{amsmath,amsthm,amscd,amsthm,upref,indentfirst}
\usepackage{amsfonts,mathrsfs}
\usepackage{amssymb,amsbsy,bm}
\usepackage{graphicx}

\usepackage{amsaddr}

\usepackage{soul}

\usepackage[us,long,24hr]{datetime}

\usepackage{mathabx}

\usepackage[usenames,dvipsnames]{color}

\theoremstyle{plain}

\theoremstyle{definition}

\theoremstyle{remark}

\numberwithin{equation}{section}
\numberwithin{theorem}{section}

\normalfont\upshape
\usepackage{exscale}
\usepackage[scaled=0.86]{helvet}

\usepackage[scr=euler,scrscaled=1.1,bb=txof,bbscaled=1.16,cal=boondox,calscaled=1.1]{mathalfa}
\renewcommand{\mathit}{\bm}
\renewcommand{\mathtt}[1]{\scalebox{1.2}{\bf \texttt{\upshape#1}}}
\renewcommand{\emph}[1]{\textcolor{blue}{\textbf{#1}}}

\usepackage{float}

\usepackage[justification=centering]{caption}

\numberwithin{equation}{section}
\numberwithin{theorem}{section}

\usepackage[normalem]{ulem}
\usepackage[square,comma]{natbib}

\usepackage{mparhack}

\usepackage{keyval}
\usepackage{totcount}
\newtotcounter{citnum} 
\def\oldbibitem{} \let\oldbibitem=\bibitem
\def\bibitem{\stepcounter{citnum}\oldbibitem}

\usepackage[verbose]{backref}
\backrefsetup{verbose=false}
\renewcommand*{\backref}[1]{}
\renewcommand*{\backrefalt}[4]{[{\tiny%
    \ifcase #1 \textsl{Not cited}%
          \or \textsl{Cited on page}~\textcolor{BrickRed}{#2}%
          \else \textsl{Cited on pages}~\textcolor{BrickRed}{#2}%
    \fi%
    }]}

\usepackage{epigraph}

\usepackage{fancyhdr}
\author{\small\scshape S\lowercase{teven} D\lowercase{uplij}}

\address{
University of M\"unster,
D-48149 M\"unster,
Germany}
\email{\small \sf douplii@uni-muenster.de;
sduplij@gmail.com;
http://www.uni-muenster.de/IT.StepanDouplii}

\title{\large\bfseries\scshape
P\lowercase{olyadic supersymmetry}}

\date{\textit{of start} April 15, 2024. \textit{Date}:
\textit{of completion}
June 3, 2024.
\newline
\mbox{}\hskip 1.16em
\textit{Total}:
24
references.
}

\renewcommand{\refname}{\textsc{References}}

\let\origsection\section
\renewcommand{\section}[1]{\sectionmark{#1}\origsection{#1}}
\let\origsubsection\subsection
\renewcommand{\subsection}[1]{\subsectionmark{#1}\origsubsection{#1}}

\makeatletter
\renewenvironment{thebibliography}[1]{%
  \@xp\origsection\@xp*\@xp{\refname}%
  \normalfont\footnotesize\labelsep .9em\relax
  \renewcommand\theenumiv{\arabic{enumiv}}\let\p@enumiv\@empty
  \vspace*{-5pt}
  \list{\@biblabel{\theenumiv}}{\settowidth\labelwidth{\@biblabel{#1}}%
    \leftmargin\labelwidth \advance\leftmargin\labelsep
    \usecounter{enumiv}}%
  \sloppy \clubpenalty\@M \widowpenalty\clubpenalty
  \sfcode`\.=\@m
}{%
  \def\@noitemerr{\@latex@warning{Empty `thebibliography' environment}}%
  \endlist
}
\makeatother

\subjclass[2010]{17A70, 17B05, 17B70, 17B80, 17B81,
 20H20, 20N10, 20N15, 81Q60, 81T60}
\keywords{superalgebra, superbracket, Lie superalgebra, sigma matrix, Pauli matrix,  arity, polyadic structure, $n$-ary group, polyadic unit, $n$-ary superalgebra, querelement, cyclic shift matrix, supersymmetry, supercharge, Hamiltonian, supersymmetric quantum mechanics}

\begin{document}
\mbox{}
\vskip 1.8cm
\begin{abstract}

\noindent We introduce a polyadic analog of supersymmetry by considering the
polyadization procedure (proposed by the author) applied to the toy model of
one-dimensional supersymmetric quantum mechanics. The supercharges are
generalized to polyadic ones using the $n$-ary sigma matrices defined in
earlier work. In this way, polyadic analogs of supercharges and Hamiltonians
take the cyclic shift block matrix form, and they can describe
multidegenerated quantum states in a way that is different from the
$N$-extended and multigraded SQM. While constructing the corresponding
supersymmetry as an $n$-ary Lie superalgebra ($n$ is the arity of the initial
associative multiplication), we have found new brackets with a reduced arity
of $2\leq m<n$ and a related series of $m$-ary superalgebras (which is
impossible for binary superalgebras). In the case of even reduced arity $m$ we
obtain a tower of higher order (as differential operators) even Hamiltonians,
while for $m$ odd we get a tower of higher order odd supercharges, and the
corresponding algebra consists of the odd sector only.

\end{abstract}

\maketitle

\thispagestyle{empty}
\mbox{}
\vspace{-0.5cm}
\tableofcontents
\newpage

\pagestyle{fancy}

\addtolength{\footskip}{15pt}

\renewcommand{\sectionmark}[1]{%
\markboth{
{ \scshape #1}}{}}

\renewcommand{\subsectionmark}[1]{%
\markright{
\mbox{\;}\\[5pt]
\textmd{#1}}{}}

\fancyhead{}
\fancyhead[EL,OR]{\leftmark}
\fancyhead[ER,OL]{\rightmark}
\fancyfoot[C]{\scshape -- \textcolor{BrickRed}{\thepage} --}

\renewcommand\headrulewidth{0.5pt}
\fancypagestyle {plain1}{ %
\fancyhf{}
\renewcommand {\headrulewidth }{0pt}
\renewcommand {\footrulewidth }{0pt}
}

\fancypagestyle{plain}{ %
\fancyhf{}
\fancyhead[C]{\scshape S\lowercase{teven} D\lowercase{uplij} \hskip 0.7cm \MakeUppercase{Polyadic Hopf algebras and quantum groups}}
\fancyfoot[C]{\scshape - \thepage  -}
\renewcommand {\headrulewidth }{0pt}
\renewcommand {\footrulewidth }{0pt}
}

\fancypagestyle{fancyref}{ %
\fancyhf{} 
\fancyhead[C]{\scshape R\lowercase{eferences} }
\fancyfoot[C]{\scshape -- \textcolor{BrickRed}{\thepage} --}
\renewcommand {\headrulewidth }{0.5pt}
\renewcommand {\footrulewidth }{0pt}
}

\fancypagestyle{emptyf}{
\fancyhead{}
\fancyfoot[C]{\scshape -- \textcolor{BrickRed}{\thepage} --}
\renewcommand{\headrulewidth}{0pt}
}
\thispagestyle{emptyf}

\section{\textsc{Introduction}}

The most natural way to introduce new symmetries and to investigate their
initial properties is by considering toy physical models. In the case of
supersymmetry, such a model is supersymmetric quantum mechanics (SQM), for a
review, see \cite{coo/kha/suk,junker,gan/bou/ras}. Its simplest
one-dimensional $N=2$ version was considered in \cite{wit2}, and then numerous
generalizations were proposed, such as, e.g., higher $N$ SQM
\cite{aku/kud,pas1986}, higher order SQM \cite{Kha1,rob/smi,fer/gar},
parasupersymmetric quantum mechanics \cite{Bec/Deb2,Rub/Spi}, and multigraded
SQM \cite{bru/dup2020sqm,top2021a,aiz/ama/shu}.

From the other side, polyadic (or higher arity) algebraic structures (for a
mathematical review, see \cite{duplij2022} and refs. therein) appeared in some
physical applications, e.g., \cite{ker2000,cast1}, including supersymmetry
\cite{bar/gun1,bag/lam1}. In addition, independently of any models, the
polyadic analog of sigma matrices ($\sigma$-matrices, or Pauli matrices) of
cyclic shift shape was proposed in \cite{dup2024p}.

In this paper we use the polyadic (nonderived $n$-ary) sigma matrices to
construct the corresponding polyadic analogs of supercharges and Hamiltonians
of cyclic shift block matrix form which can describe multi-degenerated quantum
states in a special way that is different from $N$-extended and multigraded
SQM. While it is common to endow the $n$-ary associative algebra of generators
with an $n$-ary Lie superbracket to obtain an $n$-ary Lie superalgebra, we
have constructed (using polyadic units) a series of additional superbrackets
having reduced arities $2\leq m<n$ and their related $m$-ary superalgebras.
The cases of even and odd reduced arities $m$ are significantly different: in
the former we have the tower of higher order (as differential operators) even
Hamiltonians, while in the latter case we get higher order odd supercharges
and no even elements at all. This shows that polyadic supersymmetry has a
non-trivially rich and complicated structure even in the simplest example of
the SQM model.

\section{\textsc{Polyadic sigma matrices}}

Polyadic sigma matrices were introduced in \cite{dup2024p} by using the
polyadization procedure proposed in \cite{duplij2022}. In explicit form the
full polyadic $\mathit{\Sigma}$-matrices over $\mathbb{C}$ of size $2\left(
n-1\right)  \times2\left(  n-1\right)  $ are%
\begin{equation}
\mathit{\Sigma}_{j}=\mathit{\Sigma}_{j}^{\left[  \mathbf{n}\right]  }=\left(
\begin{array}
[c]{ccccccc}%
\mathsf{0} & \fbox{$\sigma_{j}$\textsf{\ \ \ }$^{\left(  \mathbf{1}\right)  }%
$} & \mathsf{0} & \ldots & \mathsf{0} & \ldots & \mathsf{0}\\
\mathsf{0} & \mathsf{0} & \fbox{$\sigma_{j}$\textsf{\ \ \ }$^{\left(
\mathbf{2}\right)  }$} & \ldots & \mathsf{0} & \ldots & \mathsf{0}\\
\vdots & \vdots & \vdots & \ddots & \vdots & \ldots & \vdots\\
\mathsf{0} & \mathsf{0} & \mathsf{0} & \ldots & \fbox{$\sigma_{j}%
$\ \ $^{\left(  \mathbf{k}\right)  }$} & \ldots & \mathsf{0}\\
\vdots & \vdots & \vdots & \vdots & \vdots & \ddots & \vdots\\
\mathsf{0} & \mathsf{0} & \mathsf{0} & \ldots & \mathsf{0} & \ldots &
\fbox{$\sigma_{j}$\textsf{\ \ \ }$^{\left(  \mathbf{n-2}\right)  }$}\\
\fbox{$\sigma_{j}$\textsf{\ \ \ }$^{\left(  \mathbf{n-1}\right)  }$} &
\mathsf{0} & \mathsf{0} & \ldots & \mathsf{0} & \ldots & \mathsf{0}%
\end{array}
\right)  ,\ \ \ \ \ j=0,1,2,3, \label{sj}%
\end{equation}
where $\sigma_{j}$ are sigma matrices (Pauli matrices with $\sigma_{0}=I_{2}$)
and $\mathsf{0}$ is the $2\times2$ zero matrix, such that $\mathit{\Sigma}%
_{j}^{\left[  \mathbf{n}\right]  }\in GL\left(  2\left(  n-1\right)
,\mathbb{C}\right)  $ and $\det\mathit{\Sigma}_{1,2,3}^{\left[  \mathbf{n}%
\right]  }=\left(  -1\right)  ^{n}$, $\det\mathit{\Sigma}_{0}^{\left[
\mathbf{n}\right]  }=1$. The set $\left\{  \mathbf{M}_{shift}\right\}  $ of
matrices of the general shape (\ref{sj}), cyclic shift block matrices, with
invertible $GL\left(  2,\mathbb{C}\right)  $ blocks, form a nonderived $n$-ary
group $\mathcal{G}_{shift}^{\left[  \mathbf{n}\right]  }$, because the
multiplication of $n$ matrices (only, and not fewer) is closed. We define%
\begin{equation}
\mathit{\mu}^{\left[  \mathbf{n}\right]  }\left[  \mathbf{M}_{1}%
,\mathbf{M}_{2}\ldots,\mathbf{M}_{n}\right]  =\mathbf{M}_{1}\cdot
\mathbf{M}_{2}\ldots\cdot\mathbf{M}_{n},\ \ \ \ \ \mathbf{M}_{j}\in
\mathcal{G}_{shift}^{\left[  \mathbf{n}\right]  }=\left\langle \left\{
\mathbf{M}_{shift}\right\}  \mid\mathit{\mu}^{\left[  \mathbf{n}\right]
},\overline{\left(  \ \right)  }\right\rangle , \label{mn}%
\end{equation}
where $\left(  \cdot\right)  $ is the ordinary matrix product, and
$\overline{\left(  \ \right)  }$ is the querelement. Moreover, the\textit{
}$n$-ary product (\ref{mn}) is defined only for the product of $r\left(
n-1\right)  +1$ matrices, where $n\geq3$ and $r\in\mathbb{N}$ is the polyadic
power. In this notation $\mathit{\mu}^{\left[  \mathbf{2}\right]  }$ coincides
with the ordinary binary multiplication, which allows an arbitrary number of
multipliers (see \cite{duplij2022} and \cite{dup2024p} for details).

In this notation $\mathit{\Sigma}_{0}=\mathit{\Sigma}_{0}^{\left[
\mathbf{n}\right]  }=\mathbf{E}$ is the polyadic unit of the group
$\mathcal{G}_{shift}^{\left[  \mathbf{n}\right]  }$ satisfying%
\begin{equation}
\mathit{\mu}^{\left[  \mathbf{n}\right]  }\left[  \mathbf{E},\mathbf{E}%
\ldots,\mathbf{M}\right]  =\mathbf{M,} \label{me}%
\end{equation}
where $\mathbf{M}$ in the l.h.s. can be on any place, and%
\begin{equation}
\mathit{\Sigma}_{0}^{\left[  \mathbf{n}\right]  }=\mathbf{E}=\left(
\begin{array}
[c]{ccccc}%
\mathsf{0} & I_{2} & \ldots & \mathsf{0} & \mathsf{0}\\
\mathsf{0} & \mathsf{0} & I_{2} & \ldots & \mathsf{0}\\
\mathsf{0} & \mathsf{0} & \ddots & \ddots & \vdots\\
\vdots & \vdots & \ddots & \mathsf{0} & I_{2}\\
I_{2} & \mathsf{0} & \ldots & \mathsf{0} & \mathsf{0}%
\end{array}
\right)  \in\mathcal{G}_{shift}^{\left[  \mathbf{n}\right]  }\neq
\mathbf{I}=\left(
\begin{array}
[c]{ccccc}%
I_{2} & \mathsf{0} & \ldots & \mathsf{0} & \mathsf{0}\\
\mathsf{0} & I_{2} & \mathsf{0} & \ldots & \mathsf{0}\\
\mathsf{0} & \mathsf{0} & \ddots & \ddots & \vdots\\
\vdots & \vdots & \ddots & I_{2} & \mathsf{0}\\
\mathsf{0} & \mathsf{0} & \ldots & \mathsf{0} & I_{2}%
\end{array}
\right)  \notin\mathcal{G}_{shift}^{\left[  \mathbf{n}\right]  }. \label{ee}%
\end{equation}

The corresponding associative (because the ordinary matrix product (\ref{mn})
is such) $n$-ary algebra $\mathcal{A}_{shift}^{\left[  \mathbf{n}\right]  }$
consists of the cyclic shift matrices of the general form (\ref{sj}) and the
$n$-ary multiplication (\ref{mn}) together with the (binary) matrix addition
and multiplication by scalars (as for ordinary matrices)%
\begin{equation}
\mathcal{A}_{shift}^{\left[  \mathbf{n}\right]  }=\left\langle \left\{
\mathbf{M}_{shift}\right\}  \mid\mathit{\mu}^{\left[  \mathbf{n}\right]
},\left(  +\right)  ,\overline{\left(  \ \right)  }\right\rangle . \label{am}%
\end{equation}

In general, only the product of $r\left(  n-1\right)  +1$ matrices, for
$r\in\mathbb{N}$, is defined. For instance, the Cayley table of the full
ternary $\mathit{\Sigma}$-matrices is (recall that multiplication of two
$\mathit{\Sigma}_{j}=\mathit{\Sigma}_{j}^{\left[  \mathbf{3}\right]  }$ is not
closed)%
\begin{align}
\mathit{\mu}^{\left[  \mathbf{3}\right]  }\left[  \mathit{\Sigma}%
_{k},\mathit{\Sigma}_{l},\mathit{\Sigma}_{m}\right]   &  =\delta
_{kl}\mathit{\Sigma}_{m}-\delta_{km}\mathit{\Sigma}_{l}+\delta_{lm}%
\mathit{\Sigma}_{k}+i\epsilon_{klm}\mathit{\Sigma}_{0},\label{skl}\\
\mathit{\mu}^{\left[  \mathbf{3}\right]  }\left[  \mathit{\Sigma}%
_{k},\mathit{\Sigma}_{l},\mathit{\Sigma}_{0}\right]   &  =\mathit{\mu
}^{\left[  \mathbf{3}\right]  }\left[  \mathit{\Sigma}_{k},\mathit{\Sigma}%
_{0},\mathit{\Sigma}_{l}\right]  =\mathit{\mu}^{\left[  \mathbf{3}\right]
}\left[  \mathit{\Sigma}_{0},\mathit{\Sigma}_{k},\mathit{\Sigma}_{l}\right]
=\delta_{kl}\mathit{\Sigma}_{0}+i\epsilon_{klm}\mathit{\Sigma}_{m}%
,\label{skl2}\\
\mathit{\mu}^{\left[  \mathbf{3}\right]  }\left[  \mathit{\Sigma}%
_{k},\mathit{\Sigma}_{0},\mathit{\Sigma}_{0}\right]   &  =\mathit{\mu
}^{\left[  \mathbf{3}\right]  }\left[  \mathit{\Sigma}_{0},\mathit{\Sigma}%
_{k},\mathit{\Sigma}_{0}\right]  =\mathit{\mu}^{\left[  \mathbf{3}\right]
}\left[  \mathit{\Sigma}_{0},\mathit{\Sigma}_{0},\mathit{\Sigma}_{k}\right]
=\mathit{\Sigma}_{k},\ \ \ k,l,m=1,2,3, \label{skl3}%
\end{align}
where we have to include products with $\mathit{\Sigma}_{0}$, because of
(\ref{ee}).

We get, for full ternary $\mathit{\Sigma}$-matrices, the ternary commutators%
\begin{align}
\left[  \mathbf{M}_{1},\mathbf{M}_{2},\mathbf{M}_{3}\right]  ^{\left[
\mathbf{3}\right]  }  &  =\mathit{\mu}^{\left[  \mathbf{3}\right]  }\left[
\mathbf{M}_{1},\mathbf{M}_{2},\mathbf{M}_{3}\right]  +\mathit{\mu}^{\left[
\mathbf{3}\right]  }\left[  \mathbf{M}_{3},\mathbf{M}_{1},\mathbf{M}%
_{2}\right]  +\mathit{\mu}^{\left[  \mathbf{3}\right]  }\left[  \mathbf{M}%
_{2},\mathbf{M}_{3},\mathbf{M}_{1}\right] \nonumber\\
&  -\mathit{\mu}^{\left[  \mathbf{3}\right]  }\left[  \mathbf{M}%
_{1},\mathbf{M}_{3},\mathbf{M}_{2}\right]  -\mathit{\mu}^{\left[
\mathbf{3}\right]  }\left[  \mathbf{M}_{3},\mathbf{M}_{2},\mathbf{M}%
_{1}\right]  -\mathit{\mu}^{\left[  \mathbf{3}\right]  }\left[  \mathbf{M}%
_{2},\mathbf{M}_{1},\mathbf{M}_{3}\right]  , \label{m1}%
\end{align}
as%
\begin{equation}
\left[  \mathit{\mu}^{\left[  \mathbf{3}\right]  }\left[  \mathit{\Sigma}%
_{k},\mathit{\Sigma}_{l},\mathit{\Sigma}_{m}\right]  \right]  ^{\left[
\mathbf{3}\right]  }=6i\epsilon_{klm}\mathit{\Sigma}_{0},\ \ \ k,l,m=1,2,3,
\end{equation}
and standard ternary anticommutators $\left\{  \_\right\}  ^{\left[
\mathbf{3}\right]  }$ (all $+$'s in the r.h.s. of (\ref{m1}))%
\begin{equation}
\left\{  \mathit{\mu}^{\left[  \mathbf{3}\right]  }\left[  \mathit{\Sigma}%
_{k},\mathit{\Sigma}_{l},\mathit{\Sigma}_{m}\right]  \right\}  ^{\left[
\mathbf{3}\right]  }=2\delta_{kl}\mathit{\Sigma}_{m}+2\delta_{km}%
\mathit{\Sigma}_{l}+2\delta_{lm}\mathit{\Sigma}_{k}. \label{sks}%
\end{equation}

Below we will omit $\mathit{\mu}^{\left[  \mathbf{n}\right]  }$, if it will be
clear from the context. Further properties of general $\mathit{\Sigma}%
$-matrices can be found in \cite{dup2024p}.

\section{\textsc{General scheme}}

One-dimensional supersymmetric quantum mechanics (SQM) is the simplest model
which has dynamical supersymmetry \cite{wit2}. This means that there exist
transformations converting \textquotedblleft bosons\textquotedblright\ and
\textquotedblleft fermions\textquotedblright, the Hamiltonian is among the
generators, and the symmetry algebra, in addition to commutators (Lie
algebra), contains anticommutators, thus becoming its graded version: a Lie
superalgebra (for a review, see \cite{kac3})).

\subsection{Standard binary SQM\label{ssec-sqm2}}

Recall, just to introduce notation and to show the relations to be generalized
polyadically, that in the binary case (where the arity of multiplication is
$n=2$), informally, the transition from the associative superalgebra
$\mathcal{A}^{\left[  \mathbf{2}\right]  }=\left\langle \mathsf{A}\mid
\mu^{\left[  \mathbf{2}\right]  }=\left(  \cdot\right)  ,s\text{-}%
assoc\right\rangle $ to the Lie superalgebra $\mathcal{A}_{sLie}^{\left[
\mathbf{2}\right]  }=\left\langle \mathsf{A}\mid\mathcal{L}^{\left[
\mathbf{2}\right]  },s\text{-}Jacobi\right\rangle $ can be done by replacing
the binary multiplication $\mu^{\left[  \mathbf{2}\right]  }$ by the binary
Lie superbracket ($\mathbb{Z}_{2}$-graded commutator)%
\begin{equation}
\mathcal{L}^{\left[  \mathbf{2}\right]  }\left[  a_{1},a_{2}\right]
=\mu^{\left[  \mathbf{2}\right]  }\left[  a_{1},a_{2}\right]  -\left(
-1\right)  ^{\pi\left(  a_{1}\right)  \pi\left(  a_{2}\right)  }\mu^{\left[
\mathbf{2}\right]  }\left[  a_{2},a_{1}\right]  ,\ \ \ \ a_{j}\in\mathsf{A},
\label{la}%
\end{equation}
and the graded associativity by the super Jacobi identity (over the same
underlying set, a graded linear vector space $\mathsf{A}=\mathsf{A}_{0}%
\oplus\mathsf{A}_{1}$), where the parity (even, odd) is $\pi\left(  a\right)
=0,1\in\mathbb{Z}_{2}$, with $a\in\mathsf{A}_{0,1}$. The superbracket
(\ref{la}) satisfies the Lie-anticommutation relation%
\begin{equation}
\mathcal{L}^{\left[  \mathbf{2}\right]  }\left[  a_{1},a_{2}\right]  =-\left(
-1\right)  ^{\pi\left(  a_{1}\right)  \pi\left(  a_{2}\right)  }%
\mathcal{L}^{\left[  \mathbf{2}\right]  }\left[  a_{2},a_{1}\right]  .
\label{laa}%
\end{equation}

In this way, the $1D$ and $N=2$ SQM (with 2 odd supercharges $Q_{1,2}$ and the
even Hamiltonian $\mathrm{H}$) has a symmetry which is defined by the
following Lie superalgebra $\mathfrak{osp}\left(  1\mid2\right)  $ relations
\begin{align}
\mathcal{L}^{\left[  \mathbf{2}\right]  }\left[  Q_{j},Q_{k}\right]   &
=\left\{  Q_{j},Q_{k}\right\}  ^{\left[  \mathbf{2}\right]  }=\left\{
Q_{j},Q_{k}\right\}  =2\delta_{jk}\mathrm{H},\label{lq1}\\
\mathcal{L}^{\left[  \mathbf{2}\right]  }\left[  \mathrm{H},Q_{k}\right]   &
=\left[  \mathrm{H},Q_{k}\right]  ^{\left[  \mathbf{2}\right]  }=\left[
\mathrm{H},Q_{k}\right]  =0,\ \ j,k=1,2, \label{lh}%
\end{align}
where $\left[  \_,\_\right]  ^{\left[  \mathbf{2}\right]  }$ and $\left\{
\_,\_\right\}  ^{\left[  \mathbf{2}\right]  }$ are the binary commutator and
anticommutator, respectively. In a matrix representation (quantization) the
supercharges and the Hamiltonian (as operators in the two-dimensional Hilbert
superspace) have the manifest form%
\begin{align}
\hat{Q}_{1}  &  =\frac{1}{\sqrt{2}}\left(  \sigma_{1}\mathrm{\hat{p}}%
+\sigma_{2}\mathrm{W}\left(  x\right)  \right)  ,\label{q1a}\\
\hat{Q}_{2}  &  =\frac{1}{\sqrt{2}}\left(  \sigma_{2}\mathrm{\hat{p}}%
-\sigma_{1}\mathrm{W}\left(  x\right)  \right)  ,\label{q2a}\\
\mathrm{\hat{H}}  &  =\mathrm{\hat{H}}_{Witten}=\frac{\sigma_{0}}{2}\left(
\mathrm{\hat{p}}^{2}+\mathrm{W}^{2}\left(  x\right)  \right)  +\frac
{\sigma_{3}}{2}\mathrm{W}^{\prime}\left(  x\right)
,\ \ \ \ \ \ \ \ \mathrm{\hat{p}}=-i\frac{d}{dx}, \label{hhw}%
\end{align}
where $\mathrm{W}\left(  x\right)  $ is a superpotential (an arbitrary even
complex analytic function of $x$), and the parities are%
\begin{equation}
\pi\left(  \hat{Q}_{1,2}\right)  =1,\pi\left(  \mathrm{\hat{H}}\right)
=0,\pi\left(  \sigma_{1,2}\right)  =1,\pi\left(  \sigma_{0,3}\right)
=0,\pi\left(  \mathrm{\hat{p}}\right)  =0. \label{pq}%
\end{equation}

The fermionic charge $\mathrm{\hat{F}}$ (an even operator) satisfies (in this
presentation)%
\begin{equation}
\left[  \mathrm{\hat{F}},\hat{Q}_{1}\right]  =i\hat{Q}_{2},\ \ \ \left[
\mathrm{\hat{F}},\hat{Q}_{2}\right]  =-i\hat{Q}_{1},\ \ \ \left[
\mathrm{\hat{F}},\mathrm{\hat{H}}\right]  =0,\ \ \ \mathrm{\hat{F}}=\frac
{1}{2}\sigma_{3}. \label{fq}%
\end{equation}

For further details, see \cite{coo/kha/suk,junker}.

\subsection{Polyadic superalgebra with reduced arity brackets}

We now propose a generalization of the supercharges (\ref{q1a})--(\ref{q2a})
by formal substitution of sigma matrices $\sigma_{j}$ by the corresponding
polyadic sigma matrices $\mathit{\Sigma}_{j}^{\left[  \mathbf{n}\right]  }$
(\ref{sj}). This means that the Hilbert superspace becomes $2\left(
n-1\right)  $-dimensional. In this way, two consequences can occur:

\begin{enumerate}
\item Mathematical: a special polyadic analog of Lie superalgebra (with
reduced arity brackets) can be defined.

\item Physical: a polyadic analog of even Hamiltonians with higher order
derivatives and higher order odd supercharges can appear, closing the algebra,
which means that the polyadic dynamics can be richer and more prosperous.
\end{enumerate}

Because polyadic sigma matrices obey a nonderived $n$-ary multiplication, any
variables constructed from them (linearly) form a nonderived $n$-ary algebra
(using polyadic distributivity). After endowing them parities similar to
$\sigma$-matrices (\ref{pq}), this algebra becomes the totally associative
$n$-ary superalgebra $\mathcal{A}^{\left[  \mathbf{n}\right]  }=\left\langle
\mathsf{A}\mid\mu^{\left[  \mathbf{n}\right]  },n\text{-}s\text{-}%
assoc\right\rangle $ satisfying $n$-ary $\mathbb{Z}_{2}$-graded commutativity%
\begin{equation}
\mu^{\left[  \mathbf{n}\right]  }\left[  a_{1},a_{2},\ldots a_{j-1}%
,a_{j},\ldots,a_{n}\right]  =\left(  -1\right)  ^{\pi\left(  a_{j-1}\right)
\pi\left(  a_{j}\right)  }\mu^{\left[  \mathbf{n}\right]  }\left[  a_{1}%
,a_{2},\ldots a_{j},a_{j-1},\ldots,a_{n}\right]  ,\ \label{ma}%
\end{equation}
where the parity is $\pi\left(  a_{j}\right)  =0,1,\ a_{j}\in\mathsf{A}_{0,1}%
$, and has total $n$-ary associativity (for further definitions and review of
polyadic structures, see \cite{duplij2022}).

To obtain the $n$-ary Lie superalgebra we exchange (as in the binary case) the
$n$-ary multiplication $\mu^{\left[  \mathbf{n}\right]  }$ for the $n$-ary Lie
superbracket ($n$-ary $\mathbb{Z}_{2}$-graded commutator) $\mathcal{L}%
^{\left[  \mathbf{n}\right]  }\left[  a_{1},a_{2},\ldots,a_{n}\right]  $ (of
the same arity $n$ as the initial multiplication), and graded associativity
for the $n$-ary super Jacobi identity (over the same underlying set, a graded
linear vector space $\mathsf{A}=\mathsf{A}_{0}\oplus\mathsf{A}_{1}$) to get%
\begin{equation}
\mathcal{A}_{sLie}^{\left[  \mathbf{n}\right]  }=\left\langle \mathsf{A}%
\mid\mathcal{L}^{\left[  \mathbf{n}\right]  },n\text{-}s\text{-}%
Jacobi\right\rangle .
\end{equation}
The $n$-ary Lie superbracket $\mathcal{L}^{\left[  \mathbf{n}\right]  }$
satisfies (on homogeneous elements) the following $n$-ary anticommutation
relation%
\begin{equation}
\mathcal{L}^{\left[  \mathbf{n}\right]  }\left[  a_{1},a_{2},\ldots
a_{j-1},a_{j},\ldots,a_{n}\right]  =-\left(  -1\right)  ^{\pi\left(
a_{j-1}\right)  \pi\left(  a_{j}\right)  }\mathcal{L}^{\left[  \mathbf{n}%
\right]  }\left[  a_{1},a_{2},\ldots a_{j},a_{j-1},\ldots,a_{n}\right]  ,
\label{ln}%
\end{equation}
and the $n$-ary super Jacobi identity (see, e.g.,
\cite{poj2003,bar/cal/kay/san}). In the limiting cases, when all arguments
$a_{j}$ are in one subspace $\mathsf{A}_{0}$ or $\mathsf{A}_{1}$, the $n$-ary
Lie superbracket becomes an $n$-ary anticommutator or the ordinary $n$-ary
commutator respectively%
\begin{equation}
\mathcal{L}^{\left[  \mathbf{n}\right]  }\left[  a_{1},a_{2},\ldots
,a_{n}\right]  =\left\{
\begin{array}
[c]{c}%
\sum_{\sigma\in S_{n}}\mu^{\left[  \mathbf{n}\right]  }\left[  a_{\sigma
\left(  1\right)  },a_{\sigma\left(  2\right)  },\ldots,a_{\sigma\left(
n\right)  }\right]  =\left\{  a_{1},a_{2},\ldots,a_{n}\right\}  ^{\left[
\mathbf{n}\right]  },\ \text{if }a_{j}\in\mathsf{A}_{1},\\
\sum_{\sigma\in S_{n}}\left(  -1\right)  ^{\pi\left(  \sigma\right)  }%
\mu^{\left[  \mathbf{n}\right]  }\left[  a_{\sigma\left(  1\right)
},a_{\sigma\left(  2\right)  },\ldots,a_{\sigma\left(  n\right)  }\right]
=\left[  a_{1},a_{2},\ldots,a_{n}\right]  ^{\left[  \mathbf{n}\right]
},\ \text{if }a_{j}\in\mathsf{A}_{0}\text{.}%
\end{array}
\right.  \label{lnm}%
\end{equation}

The general case was given in \cite{bar/cal/kay/san} and is too cumbersome to
be presented here, we therefore restrict consideration of its manifest form to
some lower arity examples below.

Let us look more closely at the definitions of the binary Lie superbracket
(\ref{la}) and $n$-ary Lie superbracket (\ref{lnm}). If the initial
associative $n$-ary superalgebra $\mathcal{A}^{\left[  \mathbf{n}\right]  }$
has a polyadic multiplicative unit $e$ (obeying (\ref{me})--(\ref{ee})), then
we can define a superbracket with a lower arity, $m$, than the initial $n$-ary
multiplication $\mu^{\left[  \mathbf{n}\right]  }$. So instead of (\ref{lnm}),
we propose using the reduced $m$-ary superbracket (which is, obviously,
non-Lie for $m\neq n$)%
\begin{align}
&  \mathcal{R}_{\left(  n\right)  }^{\left[  \mathbf{m}\right]  }\left[
a_{1},a_{2},\ldots,a_{m}\right] \nonumber\\
&  =\left\{
\begin{array}
[c]{c}%
\sum_{\sigma\in S_{n}}\mu^{\left[  \mathbf{n}\right]  }\left[  a_{\sigma
\left(  1\right)  },a_{\sigma\left(  2\right)  },\ldots,a_{\sigma\left(
m\right)  },\overset{n-m}{\overbrace{e,\ldots,e}}\right]  =\left\{
a_{1},a_{2},\ldots,a_{n}\right\}  _{\left(  n\right)  }^{\left[
\mathbf{m}\right]  },\ \text{if }a_{j}\in\mathsf{A}_{1},\\
\sum_{\sigma\in S_{n}}\left(  -1\right)  ^{\pi\left(  \sigma\right)  }%
\mu^{\left[  \mathbf{n}\right]  }\left[  a_{\sigma\left(  1\right)
},a_{\sigma\left(  2\right)  },\ldots,a_{\sigma\left(  m\right)  }%
,\overset{n-m}{\overbrace{e,\ldots,e}}\right]  =\left[  a_{1},a_{2}%
,\ldots,a_{n}\right]  _{\left(  n\right)  }^{\left[  \mathbf{m}\right]
},\ \text{if }a_{j}\in\mathsf{A}_{0}\text{.}%
\end{array}
\right.  \label{lmn1}%
\end{align}
where $2\leq m\leq n$, and $2\leq j\leq m$. The reduced $m$-ary superbracket
satisfies the $m$-ary anticommutation%
\begin{equation}
\mathcal{R}_{\left(  n\right)  }^{\left[  \mathbf{m}\right]  }\left[
a_{1},a_{2},\ldots a_{j-1},a_{j},\ldots,a_{m}\right]  =-\left(  -1\right)
^{\pi\left(  a_{j-1}\right)  \pi\left(  a_{j}\right)  }\mathcal{R}_{\left(
n\right)  }^{\left[  \mathbf{m}\right]  }\left[  a_{1},a_{2},\ldots
a_{j},a_{j-1},\ldots,a_{m}\right]  , \label{lla}%
\end{equation}
and, possibly, the reduced $m$-ary super Jacobi identity. In the limiting case
of equal arities $m=n$ the reduced superbracket coincides with the $n$-ary Lie
superbracket%
\begin{equation}
\mathcal{R}_{\left(  n\right)  }^{\left[  \mathbf{n}\right]  }=\mathcal{L}%
^{\left[  \mathbf{n}\right]  }. \label{rl}%
\end{equation}

In the simplest case $m=2$, the reduced binary superbracket becomes (cf.
(\ref{la}))%
\begin{equation}
\mathcal{R}_{\left(  n\right)  }^{\left[  \mathbf{2}\right]  }\left[
a_{1},a_{2}\right]  =\left\{
\begin{array}
[c]{c}%
\mu^{\left[  \mathbf{n}\right]  }\left[  a_{1},a_{2},\overset{n-2}%
{\overbrace{e,\ldots,e}}\right]  +\mu^{\left[  \mathbf{n}\right]  }\left[
a_{2},a_{1},\overset{n-2}{\overbrace{e,\ldots,e}}\right]  =\left\{
a_{1},a_{2}\right\}  _{(n)}^{\left[  \mathbf{2}\right]  },\ \ \ \text{if
}a_{1},a_{2}\in\mathsf{A}_{1},\\
\mu^{\left[  \mathbf{n}\right]  }\left[  a_{1},a_{2},\overset{n-2}%
{\overbrace{e,\ldots,e}}\right]  -\mu^{\left[  \mathbf{n}\right]  }\left[
a_{2},a_{1},\overset{n-2}{\overbrace{e,\ldots,e}}\right]  =\left[  a_{1}%
,a_{2}\right]  _{(n)}^{\left[  \mathbf{2}\right]  },\ \ \ \text{in other
cases,}%
\end{array}
\right.  \label{lca}%
\end{equation}
which may be called the reduced binary anticommutator and the reduced binary
commutator, respectively.

In this way, the polyadic superalgebra with reduced arity bracket (having
$m$-ary multiplication, while constructed from an $n$-ary associative
superalgebra with the polyadic unit $e$) becomes%
\begin{equation}
\mathcal{A}_{s\text{\textit{Red}}\left(  n\right)  }^{\left[  \mathbf{m}%
\right]  }=\left\langle \mathsf{A}\mid\mathcal{R}_{\left(  n\right)
}^{\left[  \mathbf{m}\right]  },e,m\text{-}s\text{-}Jacobi\right\rangle .
\label{ams}%
\end{equation}
If the reduced $m$-ary superbracket $\mathcal{R}_{\left(  n\right)  }^{\left[
\mathbf{m}\right]  }$ satisfies the standard $m$-ary super Jacobi identity
\cite{poj2003,bar/cal/kay/san}, then we can call $\mathcal{A}%
_{s\text{\textit{Red}}\left(  n\right)  }^{\left[  \mathbf{m}\right]  }$
(\ref{ams}) a $m$-ary polyadic analog of the $n$-ary Lie superalgebra.

Thus, from each associative $n$-ary superalgebra $\mathcal{A}^{\left[
\mathbf{n}\right]  }$ (by substituting $\mu^{\left[  \mathbf{n}\right]
}\rightarrow\mathcal{R}_{\left(  n\right)  }^{\left[  \mathbf{m}\right]  }$
and $n$-ary associativity condition of $\mathcal{A}^{\left[  \mathbf{n}%
\right]  }$ with the lower arity analog of $m$-ary super Jacobi identity, if
it exists) we can build not only a superalgebra $\mathcal{A}%
_{s\text{\textit{Red}}\left(  n\right)  }^{\left[  \mathbf{2}\right]  }$ with
the reduced arity binary bracket (which does not coincide with the ordinary
binary Lie superalgebra), but also a series of $n-2$ superalgebras
$\mathcal{A}_{s\text{\textit{Red}}\left(  n\right)  }^{\left[  \mathbf{m}%
\right]  }$ having the $m$-ary reduced superbrackets $\mathcal{R}_{\left(
n\right)  }^{\left[  \mathbf{m}\right]  }$ with arities $2\leq m\leq n$ (with
$\mathcal{R}_{\left(  n\right)  }^{\left[  \mathbf{m}\right]  }=\mathcal{L}%
^{\left[  \mathbf{n}\right]  }$ if $m=n$, see (\ref{rl})).

\subsection{Algebras of polyadic supercharges}

Let us introduce polyadic ($n$-ary) supercharges $\mathit{\hat{Q}}_{1,2}$ in
the most general form (by analogy with their binary versions (\ref{q1a}%
)--(\ref{q2a}))%
\begin{align}
\mathit{\hat{Q}}_{1}  &  =\frac{1}{\sqrt{2}}\left(  \mathit{\Sigma}%
_{1}\mathrm{\hat{p}}+\mathit{\Sigma}_{2}\mathrm{W}\left(  x\right)  \right)
,\label{q1}\\
\mathit{\hat{Q}}_{2}  &  =\frac{1}{\sqrt{2}}\left(  \mathit{\Sigma}%
_{2}\mathrm{\hat{p}}-\mathit{\Sigma}_{1}\mathrm{W}\left(  x\right)  \right)  ,
\label{q2}%
\end{align}
where $\mathrm{W}\left(  x\right)  ,$are even complex analytic functions.
Regarding parities we agree that (cf. (\ref{pq}))%
\begin{equation}
\pi\left(  \mathit{\hat{Q}}_{1,2}\right)  =1,\ \ \ \pi\left(  \mathit{\Sigma
}_{1,2}\right)  =1,\ \ \ \pi\left(  \mathit{\Sigma}_{0,3}\right)
=0,\ \ \ \pi\left(  \mathbf{E}\right)  =0. \label{pq1}%
\end{equation}

In these parity prescriptions the set of matrix operators $\left\{
\mathbf{\hat{M}}_{shift}\right\}  $ of the cycled shift form (\ref{sj})
generated by the odd polyadic supercharges $\mathit{\hat{Q}}_{1,2}$
(\ref{q1})--(\ref{q2}) with respect to the matrix operator multiplication
$\mathit{\hat{\mu}}^{\left[  \mathbf{n}\right]  }$ (\ref{am}) becomes the
noncommutative, nonassociative $n$-ary graded algebra of operators
$\widehat{\mathcal{A}}_{shift}^{\left[  \mathbf{n}\right]  }=\left\langle
\left\{  \mathbf{\hat{M}}_{shift}\right\}  \mid\mathit{\hat{\mu}}^{\left[
\mathbf{n}\right]  },\mathbf{E}\right\rangle $ (cf. (\ref{am})), where
$\mathbf{E}$ is the polyadic unit (\ref{ee}). We endow the $n$-ary
superalgebra $\widehat{\mathcal{A}}_{shift}^{\left[  \mathbf{n}\right]  }$
with the reduced arity $m$-ary superbracket $\mathit{\hat{\mu}}^{\left[
\mathbf{n}\right]  }\longrightarrow\mathcal{\hat{R}}_{\left(  n\right)
}^{\left[  \mathbf{m}\right]  }$ (\ref{lmn1}), $2\leq m\leq n$, which is
polyadic anticommutative%
\begin{align}
&  \mathcal{\hat{R}}_{\left(  n\right)  }^{\left[  \mathbf{m}\right]  }\left[
\mathbf{\hat{M}}_{1},\mathbf{\hat{M}}_{2},\ldots\mathbf{\hat{M}}%
_{j-1},\mathbf{\hat{M}}_{j},\ldots,\mathbf{\hat{M}}_{m}\right]
\nonumber\\[1pt]
&  =-\left(  -1\right)  ^{\pi\left(  \mathbf{\hat{M}}_{j-1}\right)  \pi\left(
\mathbf{\hat{M}}_{j}\right)  }\mathcal{\hat{R}}_{\left(  n\right)  }^{\left[
\mathbf{m}\right]  }\left[  \mathbf{\hat{M}}_{1},\mathbf{\hat{M}}_{2}%
,\ldots\mathbf{\hat{M}}_{j},\mathbf{\hat{M}}_{j-1},\ldots,\mathbf{\hat{M}}%
_{m}\right]  , \label{ll}%
\end{align}
and can satisfy the $m$-ary super Jacobi identity
\cite{poj2003,bar/cal/kay/san}. Thus, by analogy with (\ref{ams}), we obtain
the $m$-ary superalgebra of operators with the reduced bracket $\mathcal{\hat
{R}}_{\left(  n\right)  }^{\left[  \mathbf{m}\right]  }$ (\ref{ll}) as%
\begin{equation}
\widehat{\mathcal{A}}_{s\text{\textit{Red}}\left(  n\right)  }^{\left[
\mathbf{m}\right]  }=\left\langle \left\{  \mathbf{\hat{M}}_{shift}\right\}
\mid\mathcal{\hat{R}}_{\left(  n\right)  }^{\left[  \mathbf{m}\right]
},\mathbf{E},m\text{-}s\text{-}Jacobi\right\rangle . \label{amm}%
\end{equation}

Because the $n$-ary supercharges $\mathbf{\hat{Q}}_{\ell_{j}}$ (\ref{q1}%
)--(\ref{q2}) are odd, the polyadic superalgebra $\widehat{\mathcal{A}%
}_{s\text{\textit{Red}}\left(  n\right)  }^{\left[  \mathbf{m}\right]  }$ has
different properties for different parities of the reduced arity $m$. Indeed:

\begin{enumerate}
\item Even reduced arity $m=2m^{\prime}\leq n$, $m^{\prime}\in\mathbb{N}$. We
construct the even elements of $\widehat{\mathcal{A}}_{s\text{\textit{Red}%
}\left(  n\right)  }^{\left[  \mathbf{m}\right]  }$ which can be treated as
higher polyadic analogs of the Hamiltonian which describe the dynamics of
$m$-ary supersymmetric quantum mechanics. The reduced arity superbracket
(\ref{ll}) of $n$-ary supercharges gives the higher Hamiltonian (tower)%
\begin{equation}
\mathbf{\hat{H}}_{\left(  n\right)  }^{\left[  \mathbf{2m}^{\prime}\right]
}=\frac{1}{\left(  2m^{\prime}\right)  !}\mathcal{\hat{R}}_{\left(  n\right)
}^{\left[  \mathbf{2m}^{\prime}\right]  }\left[  \overset{2m^{\prime}%
}{\overbrace{\mathbf{\hat{Q}}_{\ell_{1}},\mathbf{\hat{Q}}_{\ell_{2}}%
,\ldots,\mathbf{\hat{Q}}_{\ell_{m}}}}\right]  ,\ \ \ \ \mathbf{\hat{Q}}%
_{\ell_{j}}\in\widehat{\mathcal{A}}_{s\text{\textit{Red}}\left(  n\right)
}^{\left[  2m^{\prime}\right]  },\ \ \ \ \ell_{j}=1,2. \label{h}%
\end{equation}
In this way, we obtain polyadic supersymmetry, because informally we have
\textquotedblleft odd\textquotedblright$^{2m^{\prime}}=$\textquotedblleft
even\textquotedblright, as the polyadic analog of ordinary binary
supersymmetry \textquotedblleft odd\textquotedblright$^{2}=$\textquotedblleft
even\textquotedblright. We can use the $n$-ary anticommutator (\ref{lnm}),
because all supercharges are odd, and add a reduced arity bracket of the
higher order Hamiltonian with polyadic supercharges as \textquotedblleft even
\textquotedblright$\bullet$\textquotedblleft odd\textquotedblright%
$^{2m^{\prime}-1}=0$, and also the polyadic analog of orthogonality of
supercharges (\ref{lq1}) with $j\neq k$. In this way we obtain an example of
polyadic supersymmetry, as $2m^{\prime}$-ary supersymmetric quantum mechanics
described by the algebra (\ref{amm}) with the $2m^{\prime}$-ary reduced
bracket (cf. the standard binary SQM (\ref{lq1})--(\ref{lh}))%
\begin{equation}
\mathbf{\hat{H}}_{\left(  n\right)  }^{\left[  \mathbf{2m}^{\prime}\right]
}=\frac{1}{\left(  2m^{\prime}\right)  !}\left\{  \overset{2m^{\prime}%
}{\overbrace{\mathbf{\hat{Q}}_{\ell_{1}},\mathbf{\hat{Q}}_{\ell_{2}}%
,\ldots,\mathbf{\hat{Q}}_{\ell_{2m^{\prime}}}},}\right\}  _{\left(  n\right)
}^{\left[  \mathbf{2m}^{\prime}\right]  },\ \ \ \ \mathbf{\hat{Q}}_{\ell_{j}%
}\in\widehat{\mathcal{A}}_{s\text{\textit{Red}}\left(  n\right)  }^{\left[
\mathbf{2m}^{\prime}\right]  },\ \ \ \ \ell_{j}=1,2. \label{hm}%
\end{equation}%
\begin{align}
\mathcal{\hat{R}}_{\left(  n\right)  }^{\left[  \mathbf{2m}^{\prime}\right]
}\left[  \mathbf{\hat{H}}_{\left(  n\right)  }^{\left[  \mathbf{2m}^{\prime
}\right]  },\overset{2m^{\prime}-1}{\overbrace{\mathbf{\hat{Q}}_{\ell_{1}%
},\mathbf{\hat{Q}}_{\ell_{2}},\ldots,\mathbf{\hat{Q}}_{\ell_{2m^{\prime}-1}}}%
}\right]   &  =0,\ \ \ \ m^{\prime}\in\mathbb{N},\label{rh}\\
\mathcal{\hat{R}}_{\left(  n\right)  }^{\left[  \mathbf{2m}^{\prime}\right]
}\left[  \mathbf{\hat{Q}}_{1},\overset{2m^{\prime}-1}{\overbrace
{\mathbf{\hat{Q}}_{2},\mathbf{\hat{Q}}_{2},\ldots,\mathbf{\hat{Q}}_{2}}%
}\right]   &  =\left\{  \mathbf{\hat{Q}}_{1},\overset{2m^{\prime}%
-1}{\overbrace{\mathbf{\hat{Q}}_{2},\mathbf{\hat{Q}}_{2},\ldots,\mathbf{\hat
{Q}}_{2}}}\right\}  _{\left(  n\right)  }^{\left[  \mathbf{2m^{\prime}%
}\right]  }=0,\label{rq1}\\
\mathcal{\hat{R}}_{\left(  n\right)  }^{\left[  \mathbf{2m^{\prime}}\right]
}\left[  \mathbf{\hat{Q}}_{2},\overset{2m^{\prime}-1}{\overbrace
{\mathbf{\hat{Q}}_{1},\mathbf{\hat{Q}}_{1},\ldots,\mathbf{\hat{Q}}_{1}}%
}\right]   &  =\left\{  \mathbf{\hat{Q}}_{2},\overset{2m^{\prime}%
-1}{\overbrace{\mathbf{\hat{Q}}_{1},\mathbf{\hat{Q}}_{1},\ldots,\mathbf{\hat
{Q}}_{1}}}\right\}  _{\left(  n\right)  }^{\left[  \mathbf{2m^{\prime}%
}\right]  }=0, \label{rq2}%
\end{align}
where $\left\{  \ \ \right\}  _{\left(  n\right)  }^{\left[  \mathbf{m}%
\right]  }$ is the $m$-ary anticommutator (\ref{lmn1}). Note that the polyadic
Hamiltonians (\ref{hm}) are invariant with respect to the interchange
$\mathbf{\hat{Q}}_{1}\leftrightarrow\mathbf{\hat{Q}}_{2}$. It is seen from
(\ref{q1})--(\ref{q2}), that the reduced arity $m$ coincides with the order of
the polyadic Hamiltonians (\ref{h}) as differential operators (higher order
SQM was considered, e.g. in \cite{fer/gar}).

\item Odd reduced arity $m=2m^{\prime}+1\geq3$, $m^{\prime}\in\mathbb{N}$. In
this case we have, informally, \textquotedblleft odd\textquotedblright%
$^{2m^{\prime}+1}=$\textquotedblleft odd\textquotedblright, and so the algebra
$\widehat{\mathcal{A}}_{s\text{\textit{Red}}\left(  n\right)  }^{\left[
\mathbf{2m}^{\prime}\mathbf{+1}\right]  }$ contains no even elements at all,
and therefore -- no supersymmetry (in its \textquotedblleft
odd\textquotedblright$^{2}=$\textquotedblleft even\textquotedblright%
\ definition). Using the reduced arity superbracket (\ref{ll}), we obtain only
the higher order supercharges which are of $\left(  2m^{\prime}+1\right)  $
order as differential operators%
\begin{equation}
\mathit{\hat{Q}}_{\left(  n\right)  }^{\left[  \mathbf{2m}^{\prime}+1\right]
}=\frac{1}{\left(  2m^{\prime}+1\right)  !}\left\{  \overset{2m^{\prime}%
+1}{\overbrace{\mathbf{\hat{Q}}_{\ell_{1}},\mathbf{\hat{Q}}_{\ell_{2}}%
,\ldots,\mathbf{\hat{Q}}_{\ell_{2m^{\prime}+1}}},}\right\}  _{\left(
n\right)  }^{\left[  \mathbf{2m}^{\prime}\mathbf{+1}\right]  },\ \mathbf{\hat
{Q}}_{\ell_{j}}\in\widehat{\mathcal{A}}_{s\text{\textit{Red}}\left(  n\right)
}^{\left[  \mathbf{2m}^{\prime}\mathbf{+1}\right]  },\ \ \ell_{j}=1,2.
\end{equation}
Note that, despite $\widehat{\mathcal{A}}_{s\text{\textit{Red}}\left(
n\right)  }^{\left[  \mathbf{2m}^{\prime}\mathbf{+1}\right]  }$ containing
only odd elements (because only odd number of multipliers are allowed to close
multiplication), it is actually a $\left(  2m^{\prime}+1\right)  $-ary
superalgebra (consisting of the odd part only) with respect to the reduced
arity superbracket which is $\left(  2m^{\prime}+1\right)  $-anticommutative
(\ref{lla}). This contrasts with the ordinary (binary) superalgebras, where
the odd part by itself is not an algebra at all, since the multiplication is
not closed.
\end{enumerate}

Note that the (classical) odd Hamiltonians were obtained by changing the even
Poisson bracket to another one, the odd Poisson bracket
\cite{vol/pas/sor/tka1}, while we have the reduced arity (quantum)
superbracket $\mathcal{\hat{R}}_{\left(  n\right)  }^{\left[  \mathbf{m}%
\right]  }$ in both ($m$ is even or odd) cases.

The fermionic charge $\mathbf{\hat{F}}$, as an even matrix operator, becomes
(cf. (\ref{fq}))%
\begin{equation}
\mathbf{\hat{F}}=\frac{1}{2}\mathit{\Sigma}_{3}.
\end{equation}

The normalization in (\ref{h}) is chosen such that all the polyadic
Hamiltonians of the binary reduced arity $m=2$ would reproduce the standard
(binary) Hamiltonian (\ref{hhw}) and the main SQM relation (\ref{lq1}). Also,
it is important to note that algebras with different reduced arity $m$ of the
superbrackets $\mathcal{\hat{R}}_{\left(  n\right)  }^{\left[  \mathbf{m}%
\right]  }$ do not intersect.

\subsection{SQM with reduced arity binary bracket}

The standard binary SQM corresponds to the case $m=n=2$, and $\mathit{\Sigma
}_{j}=\sigma_{j}$ (see \textbf{Subsection} \ref{ssec-sqm2}). If $n\geq3$ and
$m=2$, we have more interesting dynamics. Indeed, the (still) binary polyadic
algebra of Hamiltonian and supercharges $\widehat{\mathcal{A}}%
_{s\text{\textit{Red}}\left(  n\right)  }^{\left[  \mathbf{2}\right]  }$ with
the reduced arity bracket becomes%
\begin{align}
&  \mathbf{\hat{H}}_{\left(  n\right)  }^{\left[  \mathbf{m}=\mathbf{2}%
\right]  }=\frac{1}{2}\left\{  \mathit{\hat{Q}}_{1,2},\mathit{\hat{Q}}%
_{1,2}\right\}  _{\left(  n\right)  }^{\left[  \mathbf{2}\right]  }=\frac
{1}{2}\mathit{\Sigma}_{0}\left(  \mathrm{\hat{p}}^{2}+\mathrm{W}^{2}\left(
x\right)  \right)  +\frac{1}{2}\mathit{\Sigma}_{3}\mathrm{W}^{\prime}\left(
x\right)  ,\label{hu}\\
&  \ \ \ \ \ \ \ \ \ \ \ \ \ \ \ \ \left\{  \mathit{\hat{Q}}_{1}%
,\mathit{\hat{Q}}_{2}\right\}  _{\left(  n\right)  }^{\left[  \mathbf{2}%
\right]  }=0,\ \ \ \ \ \ \ \ \ \ \ \ \ \ \label{qq}\\
&  \ \ \ \ \ \ \ \ \ \ \ \ \ \ \ \ \left[  \mathbf{\hat{H}}_{\left(  2\right)
}^{\left[  \mathbf{2}\right]  },\mathit{\hat{Q}}_{1}\right]  _{\left(
n\right)  }^{\left[  \mathbf{2}\right]  }=\left[  \mathbf{\hat{H}}_{\left(
2\right)  }^{\left[  \mathbf{2}\right]  },\mathit{\hat{Q}}_{2}\right]
_{\left(  n\right)  }^{\left[  \mathbf{2}\right]  }=0. \label{hq}%
\end{align}

Thus, using (\ref{sj}), (\ref{q1})--(\ref{q2}), (\ref{h}) and (\ref{hhw}), for
the case $=1$, we obtain the polyadic SQM Hamiltonian of the reduced arity
$m=2$ in the matrix form%
\begin{equation}
\mathbf{\hat{H}}_{\left(  n\right)  }^{\left[  \mathbf{m}=\mathbf{2}\right]
}=\left(
\begin{array}
[c]{ccccccc}%
\mathsf{0} & \fbox{$\mathrm{\hat{H}}_{Witten}$\textsf{\ }$^{\left(
\mathbf{1}\right)  }$} & \mathsf{0} & \ldots & \mathsf{0} & \ldots &
\mathsf{0}\\
\mathsf{0} & \mathsf{0} & \fbox{$\mathrm{\hat{H}}_{Witten}$\textsf{\ }%
$^{\left(  \mathbf{2}\right)  }$} & \ldots & \mathsf{0} & \ldots &
\mathsf{0}\\
\mathsf{0} & \mathsf{0} & \mathsf{0} & \ddots & \vdots & \ldots & \vdots\\
\mathsf{0} & \mathsf{0} & \mathsf{0} & \ldots & \fbox{$\mathrm{\hat{H}%
}_{Witten}$\ $^{\left(  \mathbf{k}\right)  }$} & \ldots & \mathsf{0}\\
\vdots & \vdots & \vdots & \vdots & \vdots & \ddots & \vdots\\
\mathsf{0} & \mathsf{0} & \mathsf{0} & \ldots & \mathsf{0} & \ldots &
\fbox{$\mathrm{\hat{H}}_{Witten}$\textsf{\ }$^{\left(  \mathbf{n-2}\right)  }%
$}\\
\fbox{$\mathrm{\hat{H}}_{Witten}$\textsf{\ }$^{\left(  \mathbf{n-1}\right)  }%
$} & \mathsf{0} & \mathsf{0} & \ldots & \mathsf{0} & \ldots & \mathsf{0}%
\end{array}
\right)  . \label{hn}%
\end{equation}

In comparison with binary SQM (\ref{hhw}), the polyadic SQM Hamiltonians with
higher reduced arity of bracket $m\geq3$ would have new features and
properties, giving novel genuine characterizations of polyadic dynamical systems.

The general structure of the polyadic higher order even Hamiltonians and the
$m$-ary higher order odd supercharges is as follows%
\begin{align}
\mathbf{\hat{H}}_{\left(  n\right)  }^{\left[  \mathbf{2m}^{\prime}\right]  }
&  =\mathrm{C}_{0}^{\left(  2m^{\prime}\right)  }\left(  x\right)
\mathit{\Sigma}_{0}^{\left[  \mathbf{n}\right]  }+\mathrm{C}_{3}^{\left(
2m^{\prime}\right)  }\left(  x\right)  \mathit{\Sigma}_{3}^{\left[
\mathbf{n}\right]  },\label{hm1}\\
\mathit{\hat{Q}}_{\left(  n\right)  ,1}^{\left[  \mathbf{2m}^{\prime
}\mathbf{+1}\right]  }  &  =\mathrm{C}_{1}^{\left(  2m^{\prime}+1\right)
}\left(  x\right)  \mathit{\Sigma}_{1}^{\left[  \mathbf{n}\right]
}+\mathrm{C}_{2}^{\left(  2m^{\prime}+1\right)  }\left(  x\right)
\mathit{\Sigma}_{2}^{\left[  \mathbf{n}\right]  },\label{hm2}\\
\mathit{\hat{Q}}_{\left(  n\right)  ,2}^{\left[  \mathbf{2m}^{\prime
}\mathbf{+1}\right]  }  &  =\mathrm{C}_{1}^{\left(  2m^{\prime}+1\right)
}\left(  x\right)  \mathit{\Sigma}_{2}^{\left[  \mathbf{n}\right]
}-\mathrm{C}_{2}^{\left(  2m^{\prime}+1\right)  }\left(  x\right)
\mathit{\Sigma}_{1}^{\left[  \mathbf{n}\right]  },\ \ \ m^{\prime}%
\in\mathbb{N}\text{,} \label{hm3}%
\end{align}
where $\mathrm{C}_{j}^{\left(  m\right)  }\left(  x\right)  $, depending on
the potential $\mathrm{W}\left(  x\right)  $, being complex analytic functions
of $x$ and derivatives of order up to $m$. The matrix block form of
(\ref{hm1})--(\ref{hm3}) coincides with (\ref{hn}), but with the equal
$2\times2$ blocks $\mathrm{C}_{0}^{\left(  2m^{\prime}\right)  }\left(
x\right)  \sigma_{0}+\mathrm{C}_{3}^{\left(  2m^{\prime}\right)  }\left(
x\right)  \sigma_{3}$ and $\mathrm{C}_{1}^{\left(  2m^{\prime}+1\right)
}\left(  x\right)  \sigma_{1}+\mathrm{C}_{2}^{\left(  2m^{\prime}+1\right)
}\left(  x\right)  \sigma_{2}$, $\mathrm{C}_{1}^{\left(  2m^{\prime}+1\right)
}\left(  x\right)  \sigma_{2}-\mathrm{C}_{2}^{\left(  2m^{\prime}+1\right)
}\left(  x\right)  \sigma_{1}$, $m^{\prime}\in\mathbb{N}$. The initial values
coincide with the standard SQM (\ref{hhw}) and (\ref{q1})--(\ref{q2}), that is
$\mathbf{\hat{H}}_{\left(  2\right)  }^{\left[  2\right]  }=\mathrm{\hat{H}%
}_{Witten}$ and $\mathit{\hat{Q}}_{\left(  2\right)  ,1}^{\left[  1\right]
}=Q_{1}$, $\mathit{\hat{Q}}_{\left(  2\right)  ,2}^{\left[  1\right]  }=Q_{2}$
(since $\mathit{\Sigma}_{j}^{\left[  \mathbf{2}\right]  }=\sigma_{j}$).

In this way we obtain the component form for the even polyadic Hamiltonian%
\begin{equation}
\mathbf{\hat{H}}_{\left(  n\right)  }^{\left[  \mathbf{2m}^{\prime}\right]
}=\left(
\begin{array}
[c]{cccccc}%
\mathsf{0} & \left(
\begin{array}
[c]{cc}%
\mathrm{\hat{H}}_{+} & 0\\
0 & \mathrm{\hat{H}}_{-}%
\end{array}
\right)  & \mathsf{0} & \mathsf{0} & \ldots & \mathsf{0}\\
\mathsf{0} & \mathsf{0} & \left(
\begin{array}
[c]{cc}%
\mathrm{\hat{H}}_{+} & 0\\
0 & \mathrm{\hat{H}}_{-}%
\end{array}
\right)  & \mathsf{0} & \ldots & \mathsf{0}\\
\mathsf{0} & \mathsf{0} & \mathsf{0} & \left(
\begin{array}
[c]{cc}%
\mathrm{\hat{H}}_{+} & 0\\
0 & \mathrm{\hat{H}}_{-}%
\end{array}
\right)  & \ldots & \mathsf{0}\\
\vdots & \vdots & \vdots & \vdots & \ddots & \vdots\\
\mathsf{0} & \mathsf{0} & \mathsf{0} & \mathsf{0} & \ldots & \left(
\begin{array}
[c]{cc}%
\mathrm{\hat{H}}_{+} & 0\\
0 & \mathrm{\hat{H}}_{-}%
\end{array}
\right) \\
\left(
\begin{array}
[c]{cc}%
\mathrm{\hat{H}}_{+} & 0\\
0 & \mathrm{\hat{H}}_{-}%
\end{array}
\right)  & \mathsf{0} & \mathsf{0} & \mathsf{0} & \ldots & \mathsf{0}%
\end{array}
\right)  , \label{h2m}%
\end{equation}
where (from (\ref{hhw}))%
\begin{align}
\mathrm{\hat{H}}_{\pm}  &  =\mathrm{C}_{0}^{\left(  2m^{\prime}\right)
}\left(  x\right)  \pm\mathrm{C}_{3}^{\left(  2m^{\prime}\right)  }\left(
x\right)  ,\\
\mathrm{C}_{0}^{\left(  2\right)  }\left(  x\right)   &  =\frac{1}{2}\left(
\mathrm{\hat{p}}^{2}+\mathrm{W}^{2}\left(  x\right)  \right)  ,\ \ \mathrm{C}%
_{3}^{\left(  2\right)  }\left(  x\right)  =\frac{1}{2}\mathrm{W}^{\prime
}\left(  x\right)  ,
\end{align}
and the higher order $\left(  2m^{\prime}+1\right)  $-ary odd supercharges%
\begin{align}
&  \mathit{\hat{Q}}_{\left(  n\right)  ,\ell}^{\left[  \mathbf{2m}^{\prime
}\mathbf{+1}\right]  }\nonumber\\
&  =\left(
\begin{array}
[c]{cccccc}%
\mathsf{0} & \left(
\begin{array}
[c]{cc}%
0 & \hat{Q}_{\ell+}\\
\hat{Q}_{\ell-} & 0
\end{array}
\right)  & \mathsf{0} & \mathsf{0} & \ldots & \mathsf{0}\\
\mathsf{0} & \mathsf{0} & \left(
\begin{array}
[c]{cc}%
0 & \hat{Q}_{\ell+}\\
\hat{Q}_{\ell-} & 0
\end{array}
\right)  & \mathsf{0} & \ldots & \mathsf{0}\\
\mathsf{0} & \mathsf{0} & \mathsf{0} & \left(
\begin{array}
[c]{cc}%
0 & \hat{Q}_{\ell+}\\
\hat{Q}_{\ell-} & 0
\end{array}
\right)  & \ldots & \mathsf{0}\\
\vdots & \vdots & \vdots & \vdots & \ddots & \vdots\\
\mathsf{0} & \mathsf{0} & \mathsf{0} & \mathsf{0} & \ldots & \left(
\begin{array}
[c]{cc}%
0 & \hat{Q}_{\ell+}\\
\hat{Q}_{\ell-} & 0
\end{array}
\right) \\
\left(
\begin{array}
[c]{cc}%
0 & \hat{Q}_{\ell+}\\
\hat{Q}_{\ell-} & 0
\end{array}
\right)  & \mathsf{0} & \mathsf{0} & \mathsf{0} & \ldots & \mathsf{0}%
\end{array}
\right)  , \label{q2m}%
\end{align}
where (from (\ref{q1a})--(\ref{q2a}))%
\begin{align}
\hat{Q}_{1\pm}  &  =\mathrm{C}_{1}^{\left(  2m^{\prime}+1\right)  }\left(
x\right)  \mp i\mathrm{C}_{2}^{\left(  2m^{\prime}+1\right)  }\left(
x\right)  ,\\
\hat{Q}_{2\pm}  &  =\mp\mathrm{C}_{1}^{\left(  2m^{\prime}+1\right)  }\left(
x\right)  -i\mathrm{C}_{2}^{\left(  2m^{\prime}+1\right)  }\left(  x\right)
,\\
\mathrm{C}_{1}^{\left(  1\right)  }\left(  x\right)   &  =\frac{1}{\sqrt{2}%
}\mathrm{\hat{p}},\ \ \mathrm{C}_{2}^{\left(  1\right)  }\left(  x\right)
=\frac{1}{\sqrt{2}}\mathrm{W}\left(  x\right)  .
\end{align}

In the examples below we present the concrete expressions for $\mathbf{\hat
{H}}_{\left(  n\right)  }^{\left[  \mathbf{m}\right]  }$ with $n=3,4$.

\section{\textsc{SQM from ternary superalgebra}}

If $n=3$, we have an even binary Hamiltonian presented by the general formulas
(\ref{hu})--(\ref{hq}) and (\ref{hn}), such that%
\begin{equation}
\mathbf{\hat{H}}_{\left(  3\right)  }^{\left[  \mathbf{2}\right]  }=\frac
{1}{2}\mathcal{\hat{R}}_{\left(  3\right)  }^{\left[  \mathbf{2}\right]
}\left[  \mathit{\hat{Q}}_{1,2},\mathit{\hat{Q}}_{1,2}\right]  _{\left(
3\right)  }^{\left[  \mathbf{2}\right]  }=\frac{1}{2}\left\{  \mathit{\hat{Q}%
}_{1,2},\mathit{\hat{Q}}_{1,2}\right\}  _{\left(  3\right)  }^{\left[
\mathbf{2}\right]  }=\left(
\begin{array}
[c]{cc}%
\mathsf{0} & \mathrm{\hat{H}}_{Witten}\\
\mathrm{\hat{H}}_{Witten} & \mathsf{0}%
\end{array}
\right)  ,
\end{equation}
and other binary SQM relations are%
\begin{align}
&  \ \ \ \ \ \ \ \ \ \ \ \ \ \ \ \ \left\{  \mathit{\hat{Q}}_{1}%
,\mathit{\hat{Q}}_{2}\right\}  _{\left(  3\right)  }^{\left[  \mathbf{2}%
\right]  }=0,\ \ \ \ \ \ \ \ \ \ \ \ \ \ \\
&  \ \ \ \ \ \ \ \ \ \ \ \ \ \ \ \ \left[  \mathbf{\hat{H}}_{\left(  3\right)
}^{\left[  \mathbf{2}\right]  },\mathit{\hat{Q}}_{1}\right]  _{\left(
3\right)  }^{\left[  \mathbf{2}\right]  }=\left[  \mathbf{\hat{H}}_{\left(
3\right)  }^{\left[  \mathbf{2}\right]  },\mathit{\hat{Q}}_{2}\right]
_{\left(  3\right)  }^{\left[  \mathbf{2}\right]  }=0,
\end{align}
where $\mathrm{\hat{H}}_{Witten}$ is given in (\ref{hhw}). In components%
\begin{equation}
\mathbf{\hat{H}}_{\left(  3\right)  }^{\left[  \mathbf{2}\right]  }=\left(
\begin{array}
[c]{cc}%
\mathsf{0} & \left(
\begin{array}
[c]{cc}%
\mathrm{\hat{H}}_{+} & 0\\
0 & \mathrm{\hat{H}}_{-}%
\end{array}
\right) \\
\left(
\begin{array}
[c]{cc}%
\mathrm{\hat{H}}_{+} & 0\\
0 & \mathrm{\hat{H}}_{-}%
\end{array}
\right)  & \mathsf{0}%
\end{array}
\right)  , \label{h32}%
\end{equation}
where $\mathrm{\hat{H}}_{\pm}=\frac{1}{2}\left(  \mathrm{\hat{p}}%
^{2}+\mathrm{W}^{2}\left(  x\right)  \right)  \pm\frac{1}{2}\mathrm{W}%
^{\prime}\left(  x\right)  $.

Next we consider the odd reduced arity $m=3$. The higher order odd
supercharges are%
\begin{align}
\mathit{\hat{Q}}_{\left(  n=3\right)  ,1}^{\left[  \mathbf{m}=\mathbf{3}%
\right]  }  &  =\frac{1}{6}\left\{  \mathit{\hat{Q}}_{1},\mathit{\hat{Q}}%
_{1},\mathit{\hat{Q}}_{1}\right\}  _{\left(  3\right)  }^{\left[
\mathbf{3}\right]  }=3\mathit{\hat{Q}}_{\left(  n=3\right)  ,21}^{\left[
\mathbf{m}=\mathbf{3}\right]  }=3\left(  \frac{1}{6}\left\{  \mathit{\hat{Q}%
}_{2},\mathit{\hat{Q}}_{1},\mathit{\hat{Q}}_{2}\right\}  _{\left(  3\right)
}^{\left[  \mathbf{3}\right]  }\right)  ,\label{q31}\\
\mathit{\hat{Q}}_{\left(  n=3\right)  ,2}^{\left[  \mathbf{m}=\mathbf{3}%
\right]  }  &  =\frac{1}{6}\left\{  \mathit{\hat{Q}}_{2},\mathit{\hat{Q}}%
_{2},\mathit{\hat{Q}}_{2}\right\}  _{\left(  3\right)  }^{\left[
\mathbf{3}\right]  }=3\mathit{\hat{Q}}_{\left(  n=3\right)  ,12}^{\left[
\mathbf{m}=\mathbf{3}\right]  }=3\left(  \frac{1}{6}\left\{  \mathit{\hat{Q}%
}_{1},\mathit{\hat{Q}}_{2},\mathit{\hat{Q}}_{1}\right\}  _{\left(  3\right)
}^{\left[  \mathbf{3}\right]  }\right)  . \label{q32}%
\end{align}

In manifest matrix form they are%
\begin{equation}
\mathit{\hat{Q}}_{\left(  3\right)  ,\ell}^{\left[  \mathbf{3}\right]
}=\left(
\begin{array}
[c]{cc}%
\mathsf{0} & \hat{Q}_{\left(  3\right)  ,\ell}^{\left[  \mathbf{3}\right]  }\\
\hat{Q}_{\left(  3\right)  ,\ell}^{\left[  \mathbf{3}\right]  } & \mathsf{0}%
\end{array}
\right)  ,
\end{equation}
where (cf. (\ref{q1})--(\ref{q2}))%
\begin{align}
\hat{Q}_{\left(  3\right)  ,1}^{\left[  \mathbf{3}\right]  }  &  =\frac
{1}{2\sqrt{2}}\left(  \sigma_{1}\left(  \mathrm{\hat{p}}^{3}+\mathrm{W}%
^{2}\left(  x\right)  \mathrm{\hat{p}}-i\mathrm{WW}^{\prime}\left(  x\right)
\right)  +\sigma_{2}\left(  \mathrm{W}\left(  x\right)  \mathrm{\hat{p}}%
^{2}-i\mathrm{W}^{\prime}\left(  x\right)  \mathrm{\hat{p}}-\mathrm{W}%
^{\prime\prime}\left(  x\right)  +\mathrm{W}^{3}\left(  x\right)  \right)
\right)  ,\\
\hat{Q}_{\left(  3\right)  ,2}^{\left[  \mathbf{3}\right]  }  &  =\frac
{1}{2\sqrt{2}}\left(  \sigma_{2}\left(  \mathrm{\hat{p}}^{3}+\mathrm{W}%
^{2}\left(  x\right)  \mathrm{\hat{p}}-i\mathrm{WW}^{\prime}\left(  x\right)
\right)  -\sigma_{1}\left(  \mathrm{W}\left(  x\right)  \mathrm{\hat{p}}%
^{2}-i\mathrm{W}^{\prime}\left(  x\right)  \mathrm{\hat{p}}-\mathrm{W}%
^{\prime\prime}\left(  x\right)  +\mathrm{W}^{3}\left(  x\right)  \right)
\right)  .
\end{align}
In components, see (\ref{q2m}), we obtain%
\begin{equation}
\mathit{\hat{Q}}_{\left(  n=3\right)  }^{\left[  \mathbf{m}=\mathbf{3}\right]
}=\left(
\begin{array}
[c]{cc}%
\mathsf{0} & \left(
\begin{array}
[c]{cc}%
0 & \hat{Q}_{+}\\
\hat{Q}_{-} & 0
\end{array}
\right) \\
\left(
\begin{array}
[c]{cc}%
0 & \hat{Q}_{+}\\
\hat{Q}_{-} & 0
\end{array}
\right)  & \mathsf{0}%
\end{array}
\right)  ,
\end{equation}
where%
\begin{equation}
\hat{Q}_{\pm}=\frac{1}{2\sqrt{2}}\left(  \mathrm{\hat{p}}^{3}+\mathrm{W}%
^{2}\left(  x\right)  \mathrm{\hat{p}}-i\mathrm{WW}^{\prime}\left(  x\right)
\pm\left(  -i\mathrm{W}\left(  x\right)  \mathrm{\hat{p}}^{2}-\mathrm{W}%
^{\prime}\left(  x\right)  \mathrm{\hat{p}}+i\mathrm{W}^{\prime\prime}\left(
x\right)  -i\mathrm{W}^{3}\left(  x\right)  \right)  \right)  .
\end{equation}

\section{\textsc{SQM from }$4$-\textsc{ary superalgebra}}

In the case $n=4$, the polyadic SQM is described by the binary Hamiltonian (of
reduced arity $m=2$) and higher order $4$-ary Hamiltonians. For the former we
have%
\[
\mathbf{\hat{H}}_{\left(  4\right)  }^{\left[  \mathbf{2}\right]
}=\mathbf{\hat{H}}_{\left(  4\right)  ,1}^{\left[  \mathbf{2}\right]  }%
=\frac{1}{2}\mathcal{\hat{R}}_{\left(  4\right)  }^{\left[  \mathbf{2}\right]
}\left[  \mathit{\hat{Q}}_{1},\mathit{\hat{Q}}_{1}\right]  _{\left(  4\right)
}^{\left[  \mathbf{2}\right]  }=\frac{1}{2}\left\{  \mathit{\hat{Q}}%
_{1},\mathit{\hat{Q}}_{1}\right\}  _{\left(  4\right)  }^{\left[
\mathbf{2}\right]  }=\mathbf{\hat{H}}_{\left(  4\right)  ,2}^{\left[
\mathbf{2}\right]  }=\frac{1}{2}\left\{  \mathit{\hat{Q}}_{2},\mathit{\hat{Q}%
}_{2}\right\}  _{\left(  4\right)  }^{\left[  \mathbf{2}\right]  }.
\]

The second order orthogonality of supercharges is (cf. (\ref{q4}))%
\begin{equation}
\left\{  \mathit{\hat{Q}}_{1},\mathit{\hat{Q}}_{2},\right\}  _{\left(
4\right)  }^{\left[  \mathbf{2}\right]  }=0.
\end{equation}

In matrix form the binary Hamiltonian is (see (\ref{hhw}) and (\ref{hn}))%
\begin{equation}
\mathbf{\hat{H}}_{\left(  4\right)  }^{\left[  \mathbf{2}\right]  }=\left(
\begin{array}
[c]{ccc}%
0 & \mathrm{\hat{H}}_{Witten} & 0\\
0 & 0 & \mathrm{\hat{H}}_{Witten}\\
\mathrm{\hat{H}}_{Witten} & 0 & 0
\end{array}
\right)  . \label{h44}%
\end{equation}

The $4$-ary Hamiltonian of $4$th order is%
\begin{align}
\mathbf{\hat{H}}_{\left(  4\right)  }^{\left[  \mathbf{4}\right]  }  &
=\mathbf{\hat{H}}_{\left(  4\right)  ,4}^{\left[  \mathbf{4}\right]  }%
=\frac{1}{24}\left\{  \mathit{\hat{Q}}_{1},\mathit{\hat{Q}}_{1},\mathit{\hat
{Q}}_{1},\mathit{\hat{Q}}_{1}\right\}  _{\left(  4\right)  }^{\left[
\mathbf{4}\right]  }=\mathbf{\hat{H}}_{\left(  4\right)  ,1}^{\left[
\mathbf{4}\right]  }=\frac{1}{24}\left\{  \mathit{\hat{Q}}_{2},\mathit{\hat
{Q}}_{2},\mathit{\hat{Q}}_{2},\mathit{\hat{Q}}_{2}\right\}  ^{\left[
\mathbf{4}\right]  }\\
&  =3\mathbf{\hat{H}}_{\left(  4\right)  ,12}^{\left[  \mathbf{4}\right]
}=3\left(  \frac{1}{24}\left\{  \mathit{\hat{Q}}_{1},\mathit{\hat{Q}}%
_{1},\mathit{\hat{Q}}_{2},\mathit{\hat{Q}}_{2}\right\}  ^{\left[
\mathbf{4}\right]  }\right)  .
\end{align}
In matrix form%
\begin{equation}
\mathbf{\hat{H}}_{\left(  4\right)  }^{\left[  \mathbf{4}\right]  }=\left(
\begin{array}
[c]{ccc}%
0 & \mathrm{\hat{H}}_{\left(  4\right)  }^{\left[  \mathbf{4}\right]  } & 0\\
0 & 0 & \mathrm{\hat{H}}_{\left(  4\right)  }^{\left[  \mathbf{4}\right]  }\\
\mathrm{\hat{H}}_{\left(  4\right)  }^{\left[  \mathbf{4}\right]  } & 0 & 0
\end{array}
\right)  ,
\end{equation}
where%
\begin{align}
\mathrm{\hat{H}}_{\left(  4\right)  }^{\left[  \mathbf{4}\right]  }  &
=\frac{\sigma_{0}}{4}\left(  \mathrm{\hat{p}}^{4}+2\mathrm{W}^{2}\left(
x\right)  \mathrm{\hat{p}}^{2}-4\mathrm{W}\left(  x\right)  \mathrm{W}%
^{\prime}\left(  x\right)  \hat{p}+\mathrm{W}^{4}\left(  x\right)
-2\mathrm{W}\left(  x\right)  \mathrm{W}^{\prime\prime}\left(  x\right)
-\mathrm{W}^{\prime2}\left(  x\right)  \right) \\
&  +\frac{\sigma_{3}}{4}\left(  2\mathrm{W}^{\prime}\left(  x\right)
\mathrm{\hat{p}}^{2}-\mathrm{W}^{\prime\prime}\left(  x\right)  \mathrm{\hat
{p}}+2\mathrm{W}^{2}\left(  x\right)  \mathrm{W}^{\prime}\left(  x\right)
-\mathrm{W}^{\prime\prime\prime}\left(  x\right)  \right)  .
\end{align}

The condition of fourth order orthogonality for the supercharges becomes%
\begin{equation}
\left\{  \mathit{\hat{Q}}_{1},\mathit{\hat{Q}}_{1},\mathit{\hat{Q}}%
_{1},\mathit{\hat{Q}}_{2}\right\}  _{\left(  4\right)  }^{\left[
\mathbf{4}\right]  }=\left\{  \mathit{\hat{Q}}_{2},\mathit{\hat{Q}}%
_{2},\mathit{\hat{Q}}_{2},\mathit{\hat{Q}}_{1}\right\}  _{\left(  4\right)
}^{\left[  \mathbf{4}\right]  }=0. \label{q4}%
\end{equation}

The the higher order odd supercharges of third reduced arity $m=3$ are%
\begin{align}
\mathit{\hat{Q}}_{\left(  4\right)  ,1}^{\left[  \mathbf{3}\right]  }  &
=\frac{1}{6}\left\{  \mathit{\hat{Q}}_{1},\mathit{\hat{Q}}_{1},\mathit{\hat
{Q}}_{1}\right\}  _{\left(  4\right)  }^{\left[  \mathbf{3}\right]
}=3\mathit{\hat{Q}}_{\left(  4\right)  ,21}^{\left[  \mathbf{3}\right]
}=3\left(  \frac{1}{6}\left\{  \mathit{\hat{Q}}_{2},\mathit{\hat{Q}}%
_{1},\mathit{\hat{Q}}_{2}\right\}  _{\left(  4\right)  }^{\left[
\mathbf{3}\right]  }\right)  ,\label{q13}\\
\mathit{\hat{Q}}_{\left(  4\right)  ,2}^{\left[  \mathbf{3}\right]  }  &
=\frac{1}{6}\left\{  \mathit{\hat{Q}}_{2},\mathit{\hat{Q}}_{2},\mathit{\hat
{Q}}_{2}\right\}  _{\left(  4\right)  }^{\left[  \mathbf{3}\right]
}=3\mathit{\hat{Q}}_{\left(  4\right)  ,12}^{\left[  \mathbf{3}\right]
}=3\left(  \frac{1}{6}\left\{  \mathit{\hat{Q}}_{1},\mathit{\hat{Q}}%
_{2},\mathit{\hat{Q}}_{1}\right\}  _{\left(  4\right)  }^{\left[
\mathbf{3}\right]  }\right)  . \label{q23}%
\end{align}

In matrix form we have (cf. (\ref{h44}))%
\begin{equation}
\mathit{\hat{Q}}_{\left(  4\right)  }^{\left[  \mathbf{3}\right]  }=\left(
\begin{array}
[c]{ccc}%
0 & \hat{Q}_{\left(  4\right)  }^{\left[  \mathbf{3}\right]  } & 0\\
0 & 0 & \hat{Q}_{\left(  4\right)  }^{\left[  \mathbf{3}\right]  }\\
\hat{Q}_{\left(  4\right)  }^{\left[  \mathbf{3}\right]  } & 0 & 0
\end{array}
\right)  . \label{q44}%
\end{equation}
The manifest form of third order supercharges $\hat{Q}_{\left(  4\right)
}^{\left[  \mathbf{3}\right]  }$ from (\ref{q44}) are%
\begin{align}
\hat{Q}_{\left(  4\right)  ,1}^{\left[  \mathbf{3}\right]  }  &  =\frac
{\sqrt{2}}{4}\left(  \sigma_{1}\left(  \mathrm{\hat{p}}^{3}+\mathrm{W}%
^{2}\left(  x\right)  \mathrm{\hat{p}}-i\mathrm{W}\left(  x\right)
\mathrm{W}^{\prime}\left(  x\right)  \right)  +\sigma_{2}\left(
\mathrm{W}\left(  x\right)  \mathrm{\hat{p}}^{2}-\mathrm{W}^{\prime}\left(
x\right)  \mathrm{\hat{p}}-\mathrm{W}^{\prime\prime}\left(  x\right)
+\mathrm{W}^{3}\left(  x\right)  \right)  \right)  ,\\
\hat{Q}_{\left(  4\right)  ,2}^{\left[  \mathbf{3}\right]  }  &  =\frac
{\sqrt{2}}{4}\left(  \sigma_{2}\left(  \mathrm{\hat{p}}^{3}+\mathrm{W}%
^{\prime2}\left(  x\right)  \mathrm{\hat{p}}-i\mathrm{W}\left(  x\right)
\mathrm{W}^{\prime}\left(  x\right)  \right)  -\sigma_{1}\left(
\mathrm{W}\left(  x\right)  \mathrm{\hat{p}}^{2}-\mathrm{W}^{\prime}\left(
x\right)  \mathrm{\hat{p}}-\mathrm{W}^{\prime\prime}\left(  x\right)
+\mathrm{W}^{3}\left(  x\right)  \right)  \right)  ,
\end{align}
which can be compared with (\ref{q1})--(\ref{q2}).

\bigskip

\textbf{Acknowledgements}. The author is deeply grateful to Mike Hewitt,
Vladimir Tkach and Raimund Vogl for useful discussions and valuable help.

\pagestyle{emptyf}
\mbox{}

\end{document}